\documentclass[12pt,preprint]{aastex}
\usepackage{psfig}
\def\HI {H\kern0.1em{\sc i}}

\newcommand{\solmass}{\mbox{M$_\odot$}}
\def\pks{PKS\,1246$-$410}
\def\Mdot{M$_\odot$ y$^{-1}$}
\def\dg{$^{\circ}$}
\begin{document}
\title{The low-power nucleus of \pks\ in the Centaurus Cluster}
\author{G. B. Taylor\altaffilmark{1,2,3}, J. S. Sanders\altaffilmark{4}, 
A. C. Fabian\altaffilmark{4}, and S. W. Allen\altaffilmark{1}}

\altaffiltext{1}{Kavli Institute of Particle Astrophysics and Cosmology,
Stanford University, Stanford, CA 94305, USA; swa@stanford.edu}
\altaffiltext{2}{National Radio Astronomy Observatory, Socorro, NM
  87801, USA}
\altaffiltext{3}{University of New Mexico, Dept. of Physics and 
Astronomy, Albuquerque, NM 87131, USA; gbtaylor@unm.edu}
\altaffiltext{4}{Institute of Astronomy, Madingley Road, Cambridge CB3 
  0HA, UK; jss, acf@ast.cam.ac.uk}


\slugcomment{As Accepted to MNRAS}

\begin{abstract}

We present Chandra, Very Large Array (VLA), and Very Long Baseline
Array (VLBA) observations of the nucleus of NGC~4696, a giant
elliptical in the Centaurus cluster of galaxies.  Like M87 in the
Virgo cluster, \pks\ in the Centaurus cluster is a nearby example of a
radio galaxy in a dense cluster environment.  In analyzing the new
X-ray data we have found a compact X-ray feature coincident with the
optical and radio core.  While nuclear emission from the X-ray source
is expected, its luminosity is low, $<10^{40}$ erg s$^{-1}$.  We
estimate the Bondi accretion radius to be 30 pc and the accretion rate
to be 0.01 \Mdot\ which under the canonical radiative efficiency of
10\% would overproduce by 3.5 orders of magnitude the radiative
luminosity.  Much of this energy can be directed into the kinetic
energy of the jet, which over time inflates the observed cavities
seen in the thermal gas.  The VLBA observations reveal a weak nucleus
and a broad, one-sided jet extending over 25 parsecs in position angle
$-$150 degrees.  This jet is deflected on the kpc-scale to a more
east-west orientation (position angle of $-$80 degrees).

\end{abstract}


\section{Introduction}

\begin{figure}[h!]
\psfig{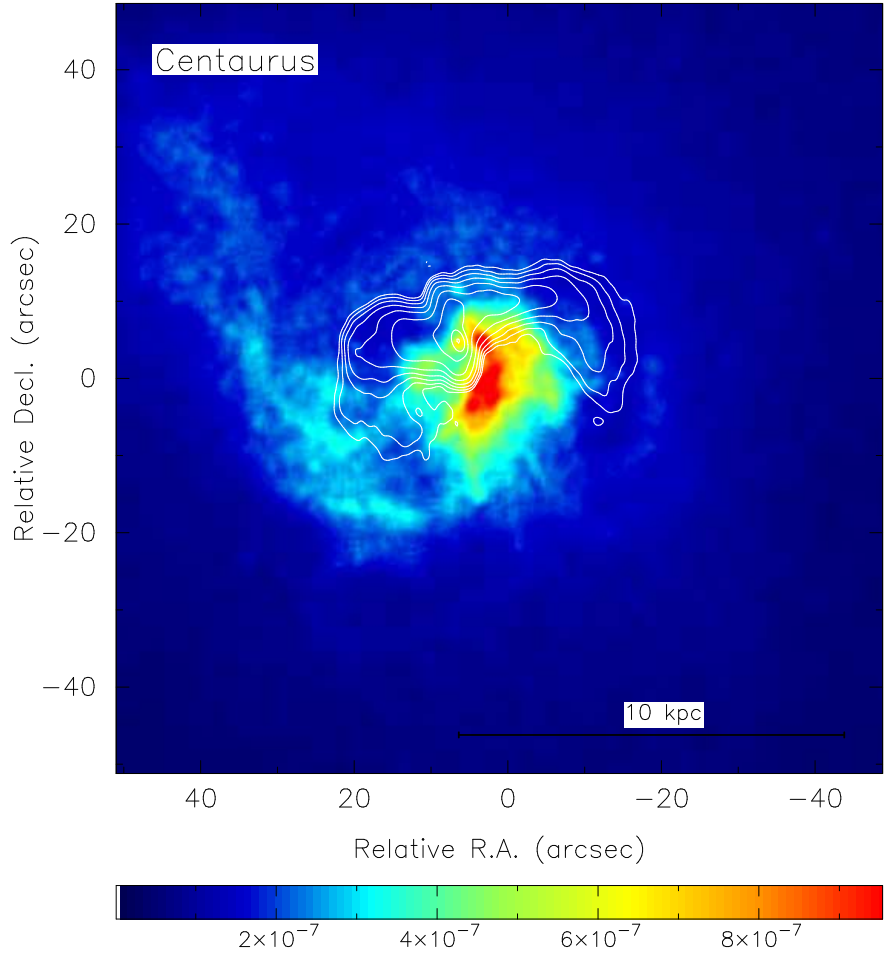}
{\caption[]{\label{cband} {\small
The central region of the Centaurus cluster in X-rays (color) and
5 GHz radio (contours).  Contour
levels begin at 0.4 mJy/beam and increase by factors of 2.  The
synthesised beam for the radio image is 2.1 $\times$ 1.2 arcseconds
in position angle 19$^\circ$. Coordinates are relative to the 
VLA pointing center at J2000 R.A. 12 48 48.7, Dec. $-$41 18 44.0.
}}}
\end{figure}

The Centaurus cluster, Abell 3526, is a nearby (redshift z=0.0104),
X-ray bright galaxy cluster.  At the center of this cluster is the
bright elliptical galaxy NGC 4696, hosting a moderately powerful radio
source, \pks.  We have been engaged in detailed studies of the X-ray
and radio emission from this cluster \citep{san02, tay02}, and have
recently obtained a further 200 ksec of Chandra data \citep{fab05}.
Our new Chandra image (Fig.~\ref{cband}), reveals a complex structure
within the central few kiloparsecs.  A plume-like structure swirls
clockwise to the NE and wraps around the radio source.  There are
clear signs of interaction between the X-ray and radio emission
including: (1) strong X-ray emission that matches the shape of the
radio emission just south of the core; (2) a faint rim of hard X-ray
emission along the northern edge of the radio source; and (3) deep
cavities in the X-ray emission on both the east and west sides.  We
also find a compact X-ray component coincident with the radio and
optical core (see Fig.~\ref{xband}).

In order to gain insights to the nature of the nucleus, we have
conducted a short, 3 hour, Very Long Baseline Array (VLBA) observation
of \pks\ at 5 GHz.  We also present a radio continuum spectrum of the
nucleus using the Very Large Array (VLA).  Both the VLA and VLBA
are operated by the NRAO\footnote{The National Radio
Astronomy Observatory is operated by Associated Universities, Inc.,
under cooperative agreement with the National Science Foundation.}.

Throughout this paper we assume $H_0$ = 70 km s$^{-1}$ Mpc$^{-1}$ so that
1\arcsec\ = 210 pc at the redshift of the Centaurus cluster.

\section{VLA Observations}\label{VLAObs}

VLA observations of the radio source \pks\ were made on
2004 October 24 at 0.326, 1.400, 4.860, 8.460, 14.940, 22.460, and
43.3140 GHz in the ``A'' configuration.  We also make use of previous
VLA observations at 4.8 and 8.3 GHz described in \cite{tay02}. Details
regarding all radio observations are given in Table~1.  The source
3C\,286 was used as the primary flux density calibrator.  Phase
calibration was derived from the nearby compact source J1316$-$3338
except for the observations at 0.326 GHz which employed a more
distant, but stronger calibrator at this frequency, J1154-3505.  These
0.326 GHz observations have been described by \cite{fab05}.
Due to poor atmospheric weather 
conditions at the VLA on 2004 October 24, phase transfer 
failed for the observations at 43 GHz.  At this high frequency,
\pks\ is too weak to self-calibrate and we were only able to obtain
an upper limit of 37 mJy on the flux density of the nucleus. 

\begin{figure}
\vspace{19cm} 
\includegraphics{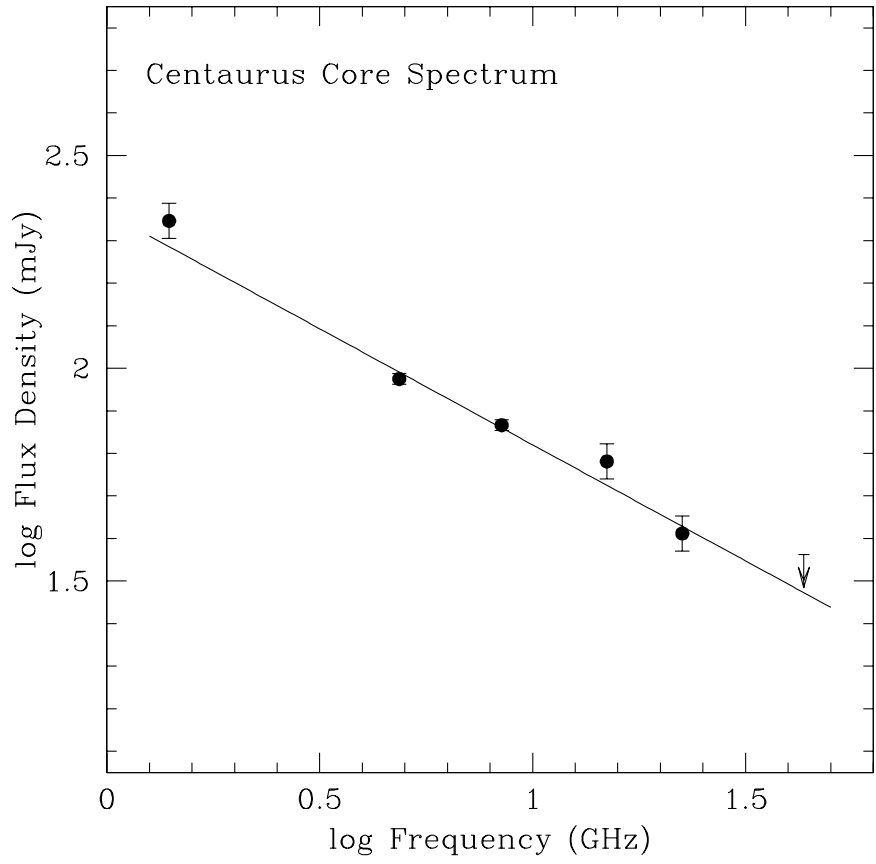}
\caption{Spectrum of the nucleus of \pks.  The solid line is a least
squares fit with slope $-0.55 \pm 0.04$. \label{spectrum}}
\end{figure}

The data were reduced in AIPS (Astronomical Image Processing System)
following the standard procedures, and imaged with Difmap.  In
Fig.~\ref{spectrum} we show a spectrum of the nucleus on 2004 Oct 24.
These data are tabulated in Table~2.  The resolution is proportional
to frequency and varies from $\sim$0.07\arcsec\ at 43 GHz to
$\sim$10\arcsec\ at 0.32 GHz.  Consequently, some lobe emission
contaminates the spectrum of the nucleus at 1.4 GHz, and we cannot
estimate the nuclear flux density at 0.32 GHz.  We find a powerlaw
with slope, $\alpha = -$0.55 $\pm$ 0.04 describes the nuclear spectrum
well between 1.4 and 43 GHz.  We define the spectral index, $\alpha$,
as $S_{\nu} \propto \nu^{\alpha}$.

\section{VLBA Observations}\label{VLBAObs}

The VLBA observed \pks\ at 5.0 GHz on 2005
February 14, with only 9 antennas since the VLBA station at Brewster,
WA, is too far north to observe this southern source.  A total of 32
MHz bandwidth was recorded in both right and left circular
polarization using 2 bit sampling.  The nearby (1.0$^\circ$) source
J1253-4059 was used for phase-referencing with a 1:1 minute cycle on
source:calibrator.  The more distant (5.1$^\circ$) calibrator
J1227-4436 was observed every 30 minutes to check on the quality of
the phase referencing and gain calibration.  Unfortunately, this source
was so distant that the phase-referencing did not succeed. 
The total time on \pks\ was 66 minutes.

Amplitude calibration was derived using measurements of the system
temperatures and antenna gains at 4992 MHz.  Fringe-fitting was
performed with the AIPS task FRING on the strong calibrator 3C~279.
Before the final application of the fringe-calibration, \pks\ was
shifted by 136 mas west, and 308 mas south based on a preliminary
measurement of its position from these observations.  The best
position of the nucleus, along with other radio properties, is
summarized in Table~3. Bandpass calibration was derived from the
observation of 3C~279.  No linear polarization calibration was
attempted for this short observing run.  Self-calibration with a 1
minute solution interval was used to further refine the calibration
and remove some slow-changing atmospheric phase errors.

\begin{figure}[h!]
\psfig{figure=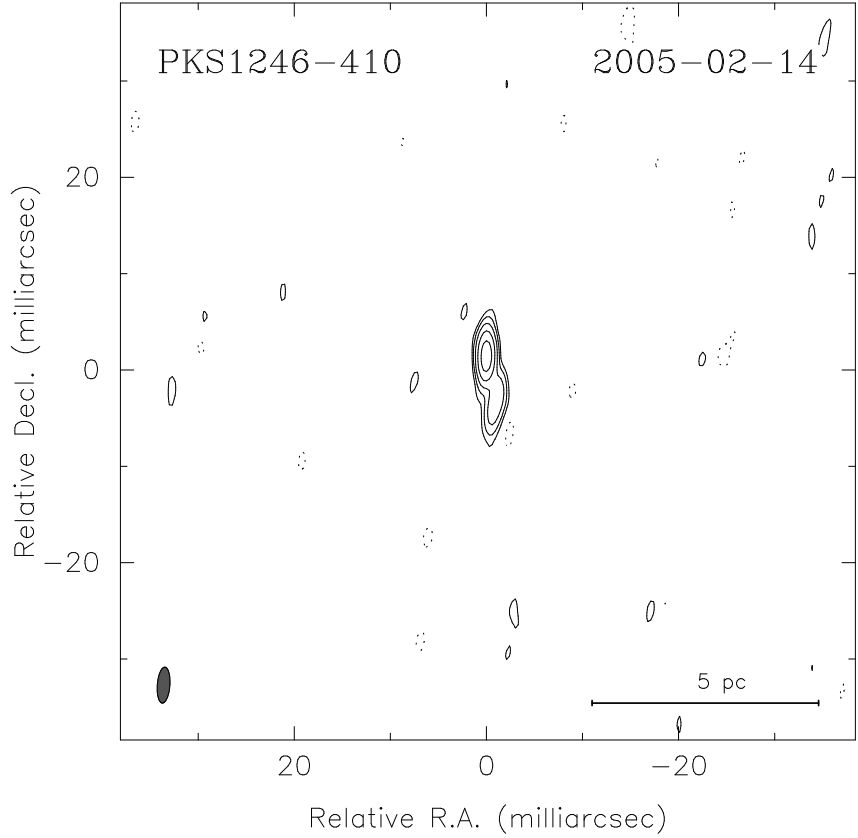,height=6.6in}
{\caption[]{\label{fullres} {\small
The full resolution image of \pks\ at 5.0 GHz.  Contours start at 
0.3 mJy/beam and increase by factors of 2.  The restoring beam of
3.74 $\times$ 1.31 mas in position angle $-$4.3$^\circ$ is drawn
in the lower left corner.   Coordinates are relative to the 
nucleus at J2000 R.A. 12 48 49.2609, Dec. $-$41 18 39.417.
}}}
\end{figure}

\begin{figure}[h!]
\psfig{figure=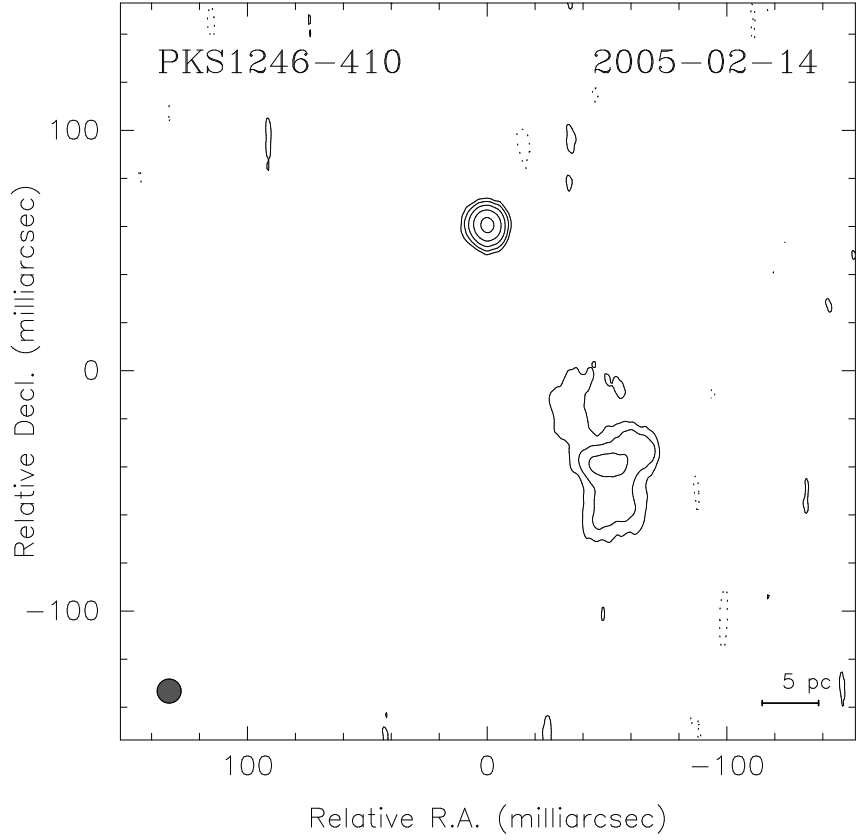,height=6.6in}
{\caption[]{\label{lowres} {\small
A tapered, low resolution image of \pks\ at 5.0 GHz.  Contours start at 
0.5 mJy/beam and increase by factors of 2.  The restoring beam of
10 mas is drawn in the lower left corner.  The zero point for the 
coordinates has been arbitrarily offset from the nucleus by 60 mas.
}}}
\end{figure}

The final image at 
full resolution (Fig.\ref{fullres}) has 
an rms noise of 90 $\mu$Jy/beam.
The full resolution VLBA image reveals a compact component with 9 mJy,
and a bright inner jet of flux density 2.4 mJy at position angle
$-$156\dg.  A tapered image (Fig.~\ref{lowres}) shows that the
jet is broad (opening angle 18\dg) and can be traced out over 120
mas in position angle $-$152\dg.  The total flux density
recovered in the VLBA image is 29 mJy.  This is 31\% of the 94.4 mJy
seen in the central 0.4\arcsec\ by the VLA, suggesting that two thirds
of the flux exists on intermediate (50 pc) scales.  The nearly
north-south alignment of the parsec-scale jet agrees well with the
initial orientation suggested by high
resolution VLA images (e.g., Fig.~\ref{cband}).  These radio properties
are similar to those of other nearby low power radio galaxies \citep{gio05}.


\section{Chandra Observations}\label{XrayObs}

The data presented here are from five Chandra observations listed in
Table~\ref{tab:xrayobs}. Each of these observations were taken using
the ACIS-S3 back-illuminated CCD. The level 1 event files were
reprocessed by CIAO 3.2 using gain file
acisD2000-01-29gain\_ctiN0003.fits. The count rate in the 2.5 to 7 keV
band on ACIS-S1 was used to filter out periods with flares using the
CIAO {\sc lc\_clean} tool, yielding a total exposure time of 196.6~ks
for the X-ray images presented here.  Each of the observations were
reprojected to have the same coordinate system as observation 4954.

For spectral analysis, we extracted spectra for a particular region
from each of the event files. Response matrices and ancillary response
matrices for each spectrum were created using the {\sc mkacisrmf}
and {\sc mkwarf} tools. Background spectra were extracted from the
same spatial regions on the chip.  These background event files were
created from blank sky observations, reprocessed and reprojected to
have the same coordinate system.  Source spectra from each run were added
together to form a total spectrum. A total weighted response and
ancillary response was created by adding together the individual
responses, weighting according to the number of counts in the 0.5 to
7~keV band of their respective foreground spectrum. The background
spectra were added, with the appropriate weight.

\begin{figure}[h!]
\psfig{figure=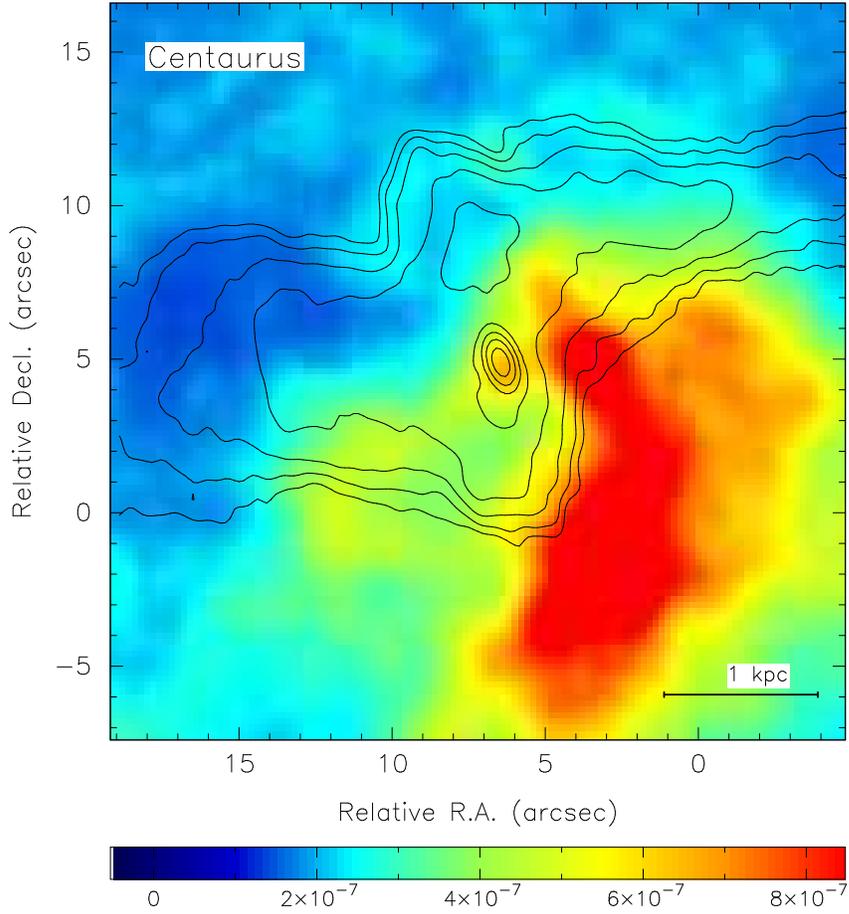,height=6.6in}
{\caption[]{\label{xband} {\small
The central region of the Centaurus cluster in X-rays (color) and
8.4 GHz radio (contours).  A weak X-ray point source is coincident with
the bright radio core.  The radio emission seems initially directed
N-S, but clearly interacts with the thermal gas and ends up
collimated E-W.  Contour
levels begin at 0.4 mJy/beam and increase by factors of 2.  
The 
synthesised beam for the radio image is 1.1 $\times$ 0.58 arcseconds
in position angle 21$^\circ$.  The zero point for the coordinates
is the same as in Fig.~1.
}}}
\end{figure}

\begin{figure}[h!]
\psfig{figure=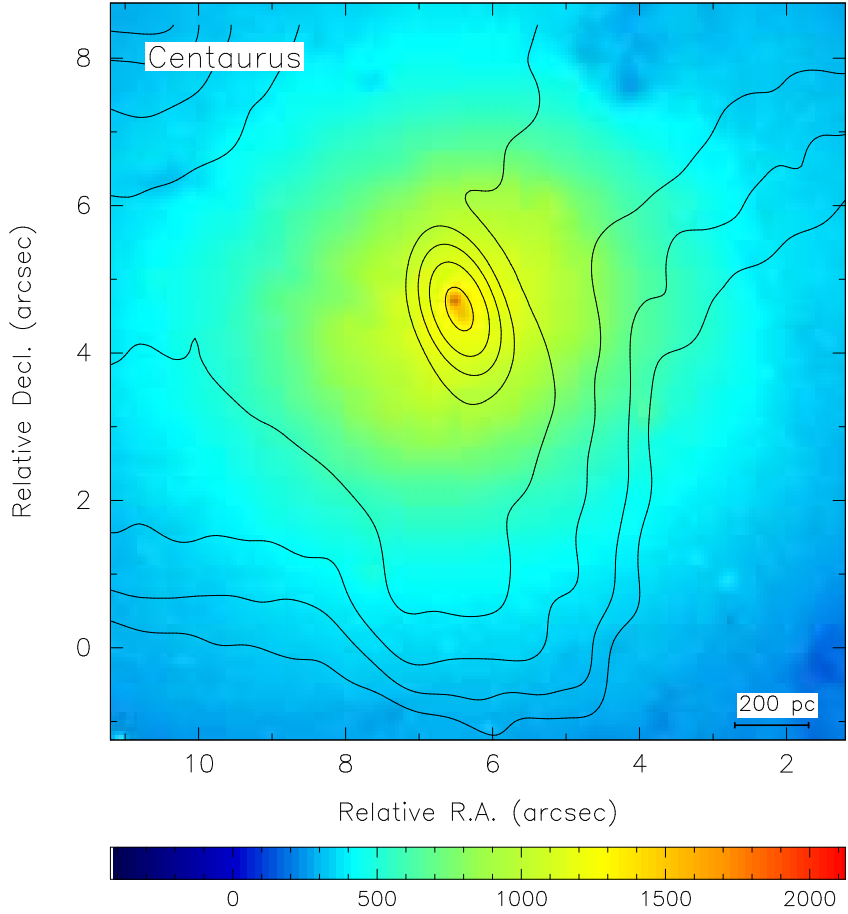,height=6.6in}
{\caption[]{\label{xopt} {\small
The central region of the Centaurus cluster from the HST at 
7904 angstroms (color) and
8.4 GHz radio (contours).  The HST image has been shifted by
0.4 arcsecond due north in order to align the centroid of the
double optical nucleus with the radio core.  The elongation of
the radio core in the same position angle as the double optical
nucleus is primarily the result of the 
synthesised beam which has dimensions 1.1 $\times$ 0.58 arcseconds
in position angle 21$^\circ$.
Contour
levels begin at 0.4 mJy/beam and increase by factors of 2.  Coordinates
are the same as Figures 1 and 2.
}}}
\end{figure}

Chandra X-ray observations of the central region (Fig.~\ref{xband})
reveal a compact source coincident with the compact radio core of
\pks.  Registration between the X-ray and radio images was tested
using observations of a serendipitous background AGN located 2
arcminutes east of \pks.  At 5 GHz a 19 mJy source is detected at RA
(J2000) 12 49 06.250 $\pm$ 0.01 and Declination (J2000) $-$41 17 50.414
$\pm$ 0.1. The source is radially smeared due to the large distance
from the center and wide bandwidth used.  The source is also detected
in the X-rays at RA (J2000) 12 49 06.231 $\pm$ 0.05 and Declination
(J2000) $-$41 17 50.086 $\pm$ 0.6.  The good agreement between the
X-ray and radio positions (Fig.~\ref{xband}) indicates that the
Chandra observations were at least as good as its nominal astrometry of
0.6 arcseconds \citep{ald00}.  This astrometry also results in good
agreement with the location of the optical nucleus (see
Fig.~\ref{xopt}).  Hubble Space Telescope (HST) observations by 
\cite{lai03} reveal a double optical nucleus with separation $\sim$0.25 
arcseconds oriented 
along a position angle of $\sim -$120 with the 
brighter component to the south-east.  The optical astrometry, however, is
not sufficiently accurate to identify which optical component is
coincident with the subparsec-scale radio core shown in Fig.~\ref{fullres}.

\begin{figure}[h!]
\psfig{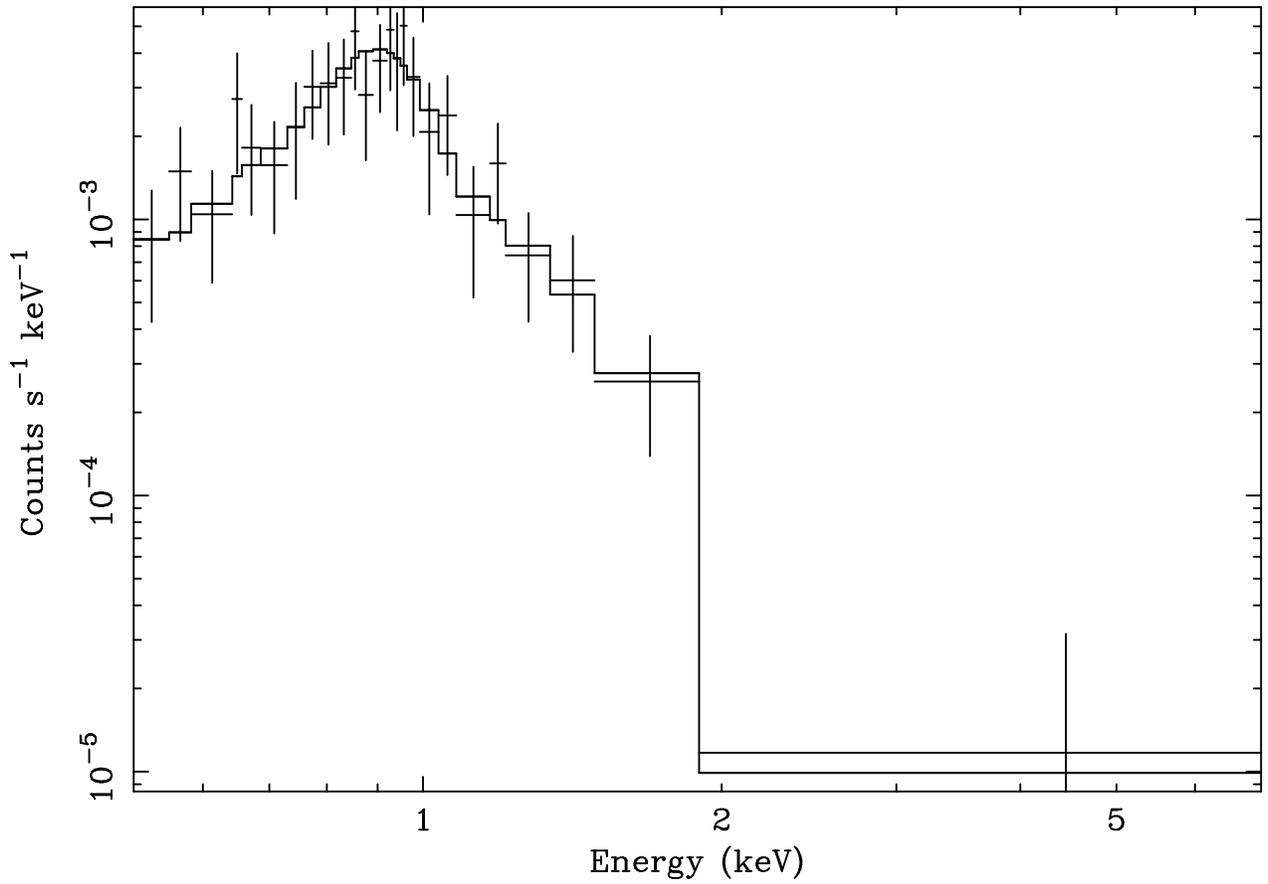}
{\caption[]{\label{xspec} {\small
The X-ray spectrum at the nucleus fit with a mekal 
spectrum, with the absorption from the Dickey \& Lockman results (plus an 
additional free absorption component at the redshift of the cluster).
A region to the east of the nucleus, 1.1 to 2.7 arcsecs in
radius, spanning 209\dg\ was used to calculate the background
level.   
}}}
\end{figure}

\begin{figure}[h!]
\psfig{figure=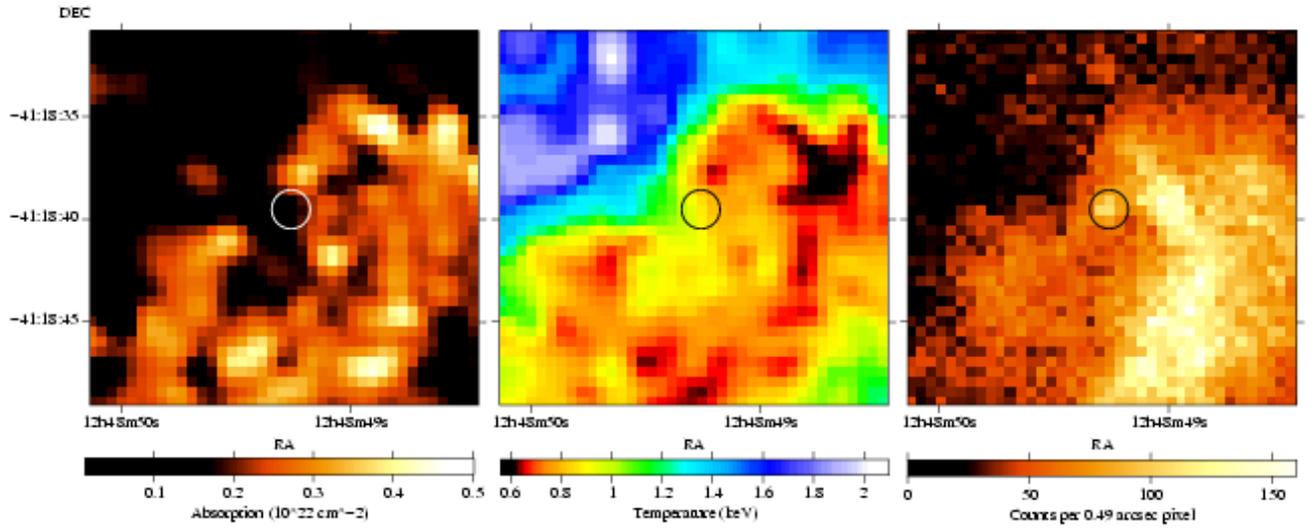,height=2.9in}
{\caption[]{\label{xtemp} {\small
The derived temperature and absorption from fits to the Chandra
X-ray observations of the inner 
part of the Centaurus cluster.  The circle indicates the location
of the nucleus.
}}}
\end{figure}

The X-ray emission from the location of the nucleus was extracted
within a circle centred on 12 48 49.26, $-$41 18 39.52 (J2000), of
radius 0.95 arcsec. The background region was a sector centred on the
same position, between radii of 1.11 and 2.67 arcsec, and at an angle
of between 6.9 and 215.9\dg\ measured from north towards the east.
This region excludes the bright thermal emission to the north and west
of the nucleus.  While compact, we cannot estimate a size for the
nucleus more precise than $\sim$0.5 arcseconds owing to the bright and
complex emission in the central region.  In Fig.~\ref{xspec} we show a
fit to the central source using a mekal spectrum, with the absorption
taken from the \cite{dic90} results.  Although an additional free
absorption component at the redshift of the cluster was allowed, the
best fit result required no additional absorption. The best fitting
parameters are a temperature of 0.75 $\pm$ 0.06 keV, and an abundance
of 0.22 $\pm$ 0.03 solar \citep{and89}. A power-law model did not
provide a good fit to the data.  The temperature distribution in the
central region of the cluster derived from the X-ray observations is
shown in Fig.~\ref{xtemp}.  To create the temperature and absorption
maps, spatial regions in the X-ray image containing approximately 400
counts between 0.5 and 7~keV were chosen using the bin accretion
tessellation algorithm of \cite{cap03}.  Spectra from these region
were independently fit using a mekal model with free absorption,
temperature, abundance and normalization in the same energy range
using C-statistics. The resulting temperature and absorption maps were
smoothed with a Gaussian of 0.49 arcseconds for display purposes. The
statistical uncertainty on the temperature is around 0.05~keV in the
central regions, and the absorption is uncertain by about $0.05 \times
10^{22}$ cm$^{-2}$.

\begin{figure}[h!]
\psfig{figure=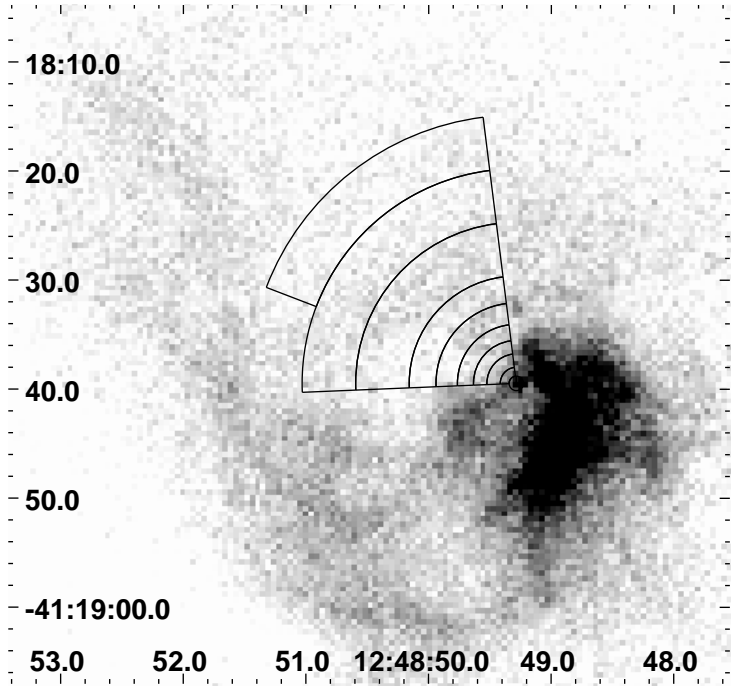,height=3.6in}
{\caption[]{\label{xtract} {\small
An unsmoothed X-ray image between 0.4 and 7 keV.  
The circular region used to extract the X-ray emission from the
compact central component is overlaid, along with the incomplete annuli used to calculate
the undisturbed radial profile.
}}}
\end{figure}

\begin{figure}[h!]
\psfig{figure=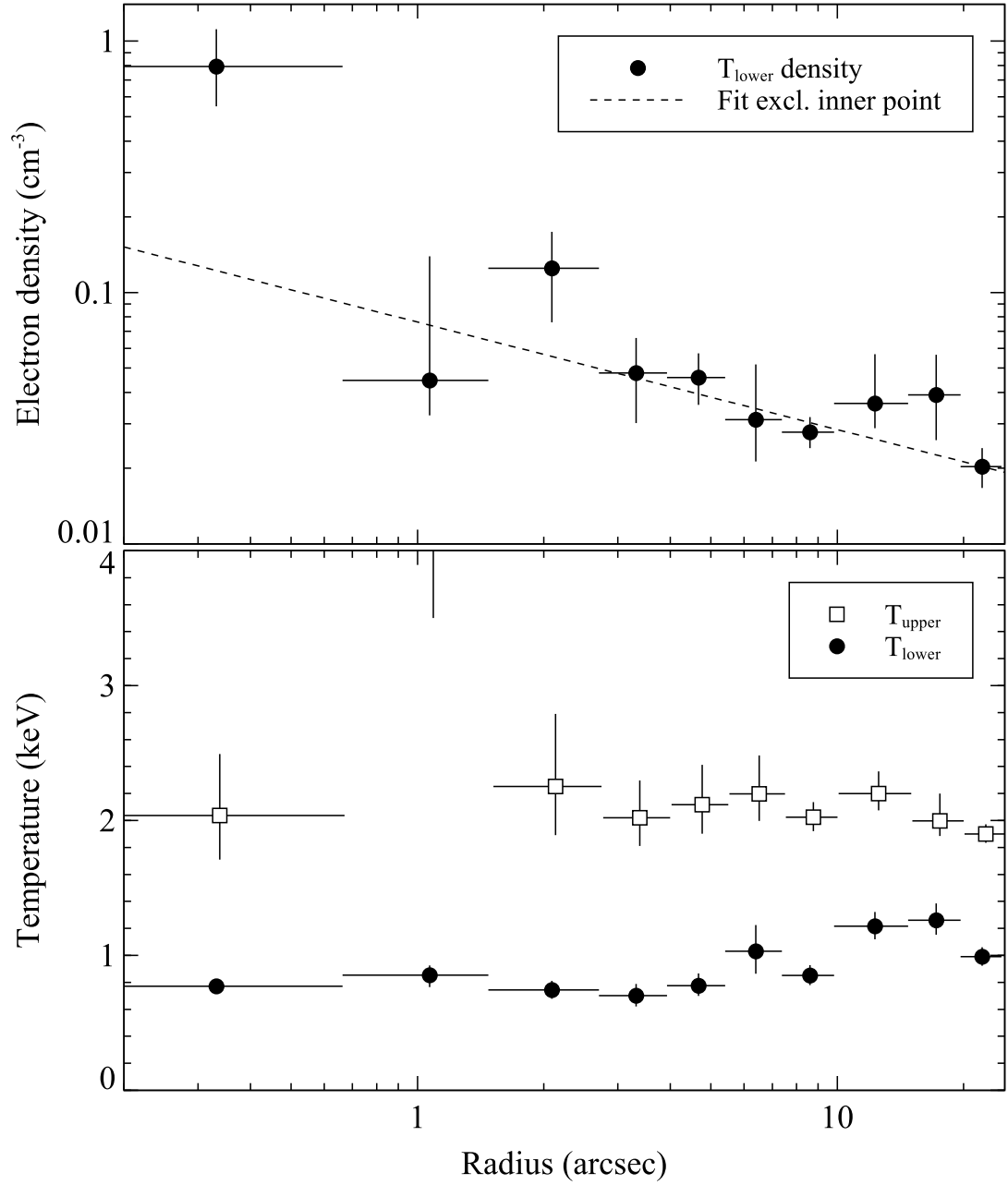,height=6.9in}
{\caption[]{\label{xfit} {\small
A fit to the temperature and density of the thermal gas in 
the central four kiloparsecs of the Centaurus cluster.
}}}
\end{figure}

To refine our estimate of the central density we fit a two-temperature
mekal model in several incomplete annuli (see Fig.~\ref{xtract}).  The
regions were chosen to avoid contamination from the bright thermal
emission to the east, and from the plume.  In this way we attempt 
to calculate the undisturbed density profile in the central region
of the Centaurus cluster (Fig.~\ref{xfit}).  By fitting a two-temperature model
this includes the effect of hotter gas projected on top of colder gas
at the center of the cluster, or any other multiphase material.
Fitting a model in which the density varies as a power law function of
the radius to all but the central point (which could be contaminated
by nuclear emission) we find an electron density:
$$
n_{\rm e} = (0.082 \pm 0.03) (r)^{-0.46 \pm 0.15} {\rm cm}^{-3}
$$
where $r$ is given in arcseconds.  

No significant change in temperature or
abundance is present at the location of the nucleus.  The luminosity
of the thermal component is 3.7 $\times 10^{39}$ erg/s between 0.1 and
10 keV. Using a powerlaw index of 1.7, and allowing for absorption
on the nucleus in addition to that on the thermal component, we derive
a 3-sigma upper limit of 1.2 $\times$ 10$^{40}$ erg/s in the 2-10 keV
band for any non-thermal power-law component (corrected for 
absorption).

\section{The Jet of \pks}\label{discussion-jet}

No counterjet is in evidence on parsec-scales, suggesting that the
nucleus is Doppler boosted.  The 3$\sigma$ limit on the counterjet
side is 0.27 mJy, providing a minimum jet:counterjet ratio of 9:1 for
the inner jet.  For the jet component at 115 mas, the surface
brightness reaches 2.7 mJy/beam in the tapered image, and a counterjet
could be detected as faint as 0.45 mJy/beam, so the lower limit on the
jet:counterjet ratio is 6:1.  Significant Doppler boosting would
produce a counterjet component closer to the nucleus, and more compact
so it is this counterpart to the larger component that we would more
naturally detect.  Assuming a spectral index of $\nu^{-0.5}$, 
we find $\beta$cos($\theta$) $>$ 0.34, implying an orientation of
less than 70\dg\ from the line-of-sight and a minimum jet 
velocity of 0.3 c (Lorentz factor, $\gamma$ $>$ 1.05).    This may indicate that
projection effects are also important on larger scales for the radio
emission, although there is some evidence that the orientation of the
jet is strongly modified on kiloparsec scales.  
It could be that dense thermal
material has stopped the expansion of the source in the north-south
direction, and that deceleration of the jet occurs on small scales in
cooling core clusters as seen in other cooling core sources
previously studied with VLBI (e.g., Hydra A, Taylor 1996 \nocite{tay96}; 
A2597, Taylor et al. 1999 \nocite{tay99}). 

Another estimate of the Lorentz factor of the jet can be obtained from
assuming a freely expanding jet (e.g., Salvati et al. 1998 \nocite{sal98}) which is
expected to show an intrinsic opening angle of 1/$\gamma$.  Under this
assumption we estimate from the observed opening angle of 18\dg\ a
Lorentz factor of 3.2.  This could be less due to projection effects
if the jet is pointed close to the line-of-sight.  A consistent
solution can be obtained with $\gamma$ = 3.2, $\beta$ = 0.95, and an
orientation of 69$^\circ$ from the line-of-sight (i.e., fairly close
to the plane of the sky).

We can estimate the minimum kinetic energy of the jet from the energy
required to evacuate the bubbles in the two lobes.  These energies are
4.5 $\times 10^{55}$ erg for the eastern bubble, and 3.7 $\times
10^{55}$ erg for the western bubble \citep{dun04}.  Assuming an age
for the bubbles of $3 \times 10^6$ years (corresponding to an average
growth rate of 0.03 c from buoyancy arguments by Dunn \& Fabian), we
get a mechanical energy input required from the jets of 8
$\times 10^{41}$ erg s$^{-1}$.  With a factor of 4
($\gamma/(\gamma-1)$ with $\gamma=4/3$) for the total energy in
relativistic particles in a bubble, together with a factor for the
lobes being overpressured since they are growing, the output could be
up to ten times larger. This falls about a factor 3 short of
the X-ray cooling luminosity,
required to stem a cooling catastrophe in a few Gyr, or about $2
\times 10^{43}$ erg s$^{-1}$.


\section{The Central Engine of \pks}\label{discussion}

In either case of an active galactic nucleus (AGN) seen directly, or
indirectly via heating, the nucleus has a relatively low power
compared to radio galaxies of similar radio luminosity
\citep{dima00,rei00,don04}.  Absorption does not seem to be important,
so the accretion onto the core must be radiatively inefficient.  The
large inclination angle of $\sim$70\dg\ (see \S 5) is consistent with
the relation between core X-ray power and inclination angle found by
\citet{don04} for a sample of FR~I radio galaxies.  \citet{don04}
suggest that the decreasing X-ray luminosity with increasing
inclination is caused by beaming of the X-ray emission.  This can be
explained if a substantial fraction of the non-thermal X-ray emission
is produced by the Doppler-boosted jet.

\cite{ber02} measure a velocity dispersion, $\sigma$, in NGC~4696 of
262 km s$^{-1}$.  Using the $M_{BH}$ - $\sigma$ relation derived by
\cite{pin03} for 10 early-type galaxies, this velocity dispersion
translates to a black hole mass, $M_{BH}$, of 
3 $\times 10^{8}$ \solmass with an uncertainty of about a factor
of 2.  This is close to that estimated
by \cite{fuj04}, based on the same measurements of $\sigma$ by
\cite{ber02} and the $M_{BH}$ - $\sigma$ relation derived by
\cite{geb00}.

The Bondi accretion radius, $r_{A}$, where the gravitational potential
dominates over the thermal energy, for a black hole of mass $M_{BH}$
is given by
$$
r_{A} =  G M_{BH} m_p / k T
$$
where $m_{p}$ is the proton mass, and $T$ is the temperature of the 
accreting material.  From this we derive a Bondi accretion radius 
of 30 pc.  At the distance of \pks\ this corresponds to a radius
of 0.14 arcsec, which is not resolved by our X-ray observations. 
From our fit to the density profile from \S 4, at 0.14 arcsec we 
estimate an electron density of 0.2 cm$^{-3}$ with an uncertainty
of a factor of 2.
The Bondi 
accretion rate within this radius is 
$$
\dot{M} = 4 \pi r_{A}^2 \rho c_s
$$
where $\rho$ is the density, and $c_s$ is the sound speed given
by $c_s$ $\sim$ 2.7 $\times 10^7 T_{0.8}^{1/2}$ cm s$^{-1}$ with
$T_{0.8} = T/0.8$ keV the ISM temperature in units of keV.  
For \pks\ we find an accretion rate of 0.014 \Mdot.  Assuming a 
10\% conversion efficiency into radiation, this predicts a 
luminosity of 8 $\times 10^{43}$ erg s$^{-1}$.  This luminosity
is nearly four orders of magnitude more than observed in X-rays (\S 4),
or to put it another way, the radiative efficiency, $\eta$, is 
$\sim2 \times 10^{-4}$.
We can also compare the accretion energy output to that of the jet (\S 5), 
and find that it is an order of magnitude larger than required
to blow the bubbles.    It could be that the Bondi accretion
has been suppressed somewhat by energy injected into the ISM by the 
jet \citep{dima03}, or by the presence of a binary black hole within
the Bondi radius, if the double optical nucleus \citep{lai03} is 
interpreted this way. Another explanation to balance the 
energy output with requirements could be that the mass
of the black hole is a factor 2 lower, and the density is similarly
overestimated, which would reduce the 
energy output by 1 order of magnitude, commensurate with the 
kinetic energy estimated for the jet.

The black hole in the nucleus of M87 has a mass of 3 $\times 10^{9}$
\solmass\ derived from HST spectroscopy \citep{for94,har94,mac97}.
The Bondi accretion rate in M87 \citep{dima03} is $\sim$0.1 \Mdot, about
eight times that in \pks\, predicting a nuclear luminosity of 
5 $\times 10^{44}$ erg s$^{-1}$.  The  
kinetic energy estimated for the jets in M87 is at least
3 $\times 10^{42}$ erg s$^{- 1}$ \citep{owe00, you02, dun04}, nearly
an order of magnitude higher than estimated for \pks.
With a factor of $\sim$10 for 
overpressured bubbles of relativistic particles, the output could be
up to ten times larger.  As for \pks, this is still about 1 order of 
magnitude less than the predicted nuclear luminosity.

Another nearby cD elliptical galaxy that has been studied with Chandra
is NGC 6166 (which hosts the radio source 3C~338) in A2199
\citep{dima01}.  In NGC 6166 the derived Bondi accretion rate is 0.03
\Mdot, providing a luminosity of 2 $\times 10^{44}$ erg s$^{-1}$ and a
similarly low X-ray luminosity and accretion efficiency is derived.
The minimum kinetic energy estimated to inflate the observed bubbles
is 5$\times 10^{55}$ erg for the eastern bubble, and 17 $\times
10^{55}$ erg for the western bubble \citep{dun04}. Again assuming
growth at the buoyant velocity, the minimum kinetic energy estimate is
6 $\times 10^{41}$ erg s$^{-1}$, but could be greater by a factor of
10 for overpressured bubbles of relativistic particles.
The jets of NGC~6166 therefore require about a factor 30 less energy than
available from Bandi accretion, somewhat less then for M87 and \pks.

\section{Conclusions}

Along with M87 \citep{dima03}, and NGC 6166 \citep{dima01}, NGC 4696
provides a third elliptical galaxy where the accretion rate onto a
supermassive black hole can be studied in detail. All three of these
systems have inner bubbles traced by the thermal gas that allow for an
estimation of the time-averaged kinetic energy of the jets
\citep{dun04}.  These energies are systematically 10-30 times lower
than the energy available from Bondi accretion (assuming 10\%
conversion efficiency).  However, this narrow range of values and the
equally remarkable narrow range in radiative efficiency for these
systems, which are all low with $\eta = 1-2 \times 10^{-5}$, may
indicate a universal property of accretion in the generation of
jets, and will be explored further in a future paper (in
preparation).  A low, but broader range of accretion efficiency of
$\eta = 10^{-2}$ to $10^{-5}$ is found by \citet{don04} for 13
FR~I galaxies from the 3CR and B2 catalogs.  

\cite{dima03} suggest that M87 may be currently in a low state.  With
the addition of \pks\ to the small number of radio galaxies (including
3C338) for which the Bondi radius can be estimated, and an accurate
value of the radiative efficiency calculated, this argument weakens.
It is more likely that for the majority of the time the accretion
efficiency is low.

We note that given its proximity and our estimated Lorentz factor of
$\sim$3, the expected motion of the jet components in \pks\ is 
2 mas yr$^{-1}$.  This motion could be readily detected in 
VLBI observations spread out over a few months.  

\acknowledgments GBT acknowledges support for this work from the
National Aeronautics and Space Administration through Chandra Award
Numbers GO4-5134X and GO4-5135X issued by the Chandra X-ray
Observatory Center, which is operated by the Smithsonian Astrophysical
Observatory for an on behalf of the National Aeronautics and Space
Administration under contract NAS8-03060.  SWA thanks the Royal
Society for support.  This research has made use of the NASA/IPAC
Extragalactic Database (NED) which is operated by the Jet Propulsion
Laboratory, Caltech, under contract with NASA.  The National Radio
Astronomy Observatory is a facility of the National Science Foundation
operated under a cooperative agreement by Associated Universities,
Inc.

\clearpage

\clearpage

\clearpage


\clearpage

\clearpage

\clearpage

\clearpage

\clearpage

\clearpage

\clearpage

\clearpage

\begin{center}
TABLE 1 \\
\smallskip
R{\sc adio} O{\sc bservational} P{\sc arameters}
\smallskip

\begin{tabular}{l l r r r r r r r r}
\hline
\hline
Source & Date & Frequency & Bandwidth & Config. & Duration \\
 &  & (MHz) & (MHz) &  & (min) \\
\hline
\noalign{\vskip2pt}
\pks\ & Oct 2004 & 323/328  & 6.25 & A & 178 \\
      & Oct 2004 & 1365/1435   & 50 & A & 1 \\
      & Apr 1998 & 4635/4885  & 50 & A & 5 \\
      & Jun 1998 & 4635/4885  & 50 & BnA & 60 \\
      & Oct 2004 & 4885/4835   & 50 & A & 1 \\
      & Feb 2005 & 4986        & 32 & VLBA & 66 \\
      & Jun 1998 & 8115/8485  & 50 & BnA & 54 \\
      & Nov 1998 & 8115/8485  & 50 & CnB & 37 \\
      & Oct 2004 & 8435/8485   & 50 & A & 1 \\
      & Oct 2004 & 14965/14915 & 50 & A & 1 \\
      & Oct 2004 & 22485/22435 & 50 & A & 2 \\
      & Oct 2004 & 43315/43365 & 50 & A & 3 \\
\hline
\end{tabular}
\end{center}
\clearpage

\begin{center}
TABLE 2 \\
\smallskip
VLA C{\sc ore} F{\sc lux} D{\sc ensity} \\
\smallskip
\begin{tabular}{l r r}
\hline
\hline
Frequency & Flux & rms \\
 (MHz) & (mJy) & (mJy/beam) \\
\hline
\noalign{\vskip2pt}
 1400  & 222 $\pm$ 22 &  0.56 \\
 4860  & 94.4 $\pm$ 2.8 & 1.3 \\
 8460  & 73.5 $\pm$ 2.2 & 0.50 \\
 14940  & 60.4 $\pm$ 6.0 & 0.73 \\
 22460  & 40.9 $\pm$ 4.1 & 0.59 \\
 43340  & $<$36.5 & 3.3 \\
\hline
\end{tabular}
\end{center}

\clearpage

\begin{center}
TABLE 3 \\
\smallskip
S{\sc ource} P{\sc roperties} \\
\smallskip
\begin{tabular}{l c c}
\hline
\hline
Property & PKS\,2322$-$123 \\
\hline
\noalign{\vskip2pt}
core RA (J2000) & 12$^h$48$^m$49\rlap{$^s$}{.\,}2609 \\ 
\phantom{core}Dec. (J2000) &  $-$41\arcdeg 18\arcmin 39\rlap{\arcsec}{.\,}417 \\
\phantom{core}Gal. lat. \phantom{   }($b$) & 21.56\dg  \\
\phantom{core}Gal. long. ($l$) & 302.40\dg \\
radial velocity & 2958 $\pm$ 15 km s$^{-1}$ \\
distance from cluster center &  0.0 Mpc \\
luminosity distance & 43.0 Mpc \\
core flux density (5 GHz) & 9.0 $\pm$ 0.5 mJy \\
core power (5 GHz) & 2.0 $\times$ 10$^{21}$ W Hz$^{-1}$ \\
largest angular size & 150\arcsec \\
largest physical size & 30 kpc \\
total flux density (5 GHz) & 1330 mJy \\
total power (5 GHz) & 3.0 $\times$ 10$^{23}$ W Hz$^{-1}$ \\
\hline
\end{tabular}
\end{center}

\clearpage

\begin{center}
TABLE 4 \\
\smallskip
C{\sc handra} E{\sc xposures} \\
\smallskip
\begin{tabular}{llllll}
\hline
\hline
OBSID  & Date  & Clean exposure & RA & DEC  & Y offset \\ 
       &       & (ksec)  & (J2000) & (J2000) & (arcmin) \\ 
(1)    & (2)   & (3)  & (4) & (5) & (6) \\ 
\hline
\noalign{\vskip2pt}
504    & 2000-05-22 & 21.8  & 12 48 48.7 & -41 18 44.0 & -1.0 \\
505    & 2000-06-08 & 10.0  & 12 48 48.7 & -41 18 44.0 & -1.0 \\
4954   & 2004-04-01 & 80.2  & 12 48 48.9 & -41 18 44.4 & -1.3 \\
4955   & 2004-04-02 & 40.6  & 12 48 48.9 & -41 18 44.4 & -0.8 \\
5310   & 2004-04-04 & 44.1  & 12 48 48.9 & -41 18 44.4 & -1.8 \\
\hline
\end{tabular}
\end{center}
{Notes: (1) Chandra Observation ID; (2) date of observation; (3) 
cleaned exposure time; (4) target RA and (5) DEC; (6) offset of the 
target from the aimpoint in the Y direction of the CCD.}
\label{tab:xrayobs}
\clearpage

\end{document}